\documentstyle[12pt]{article}
\begin{document}
\title{The Fractal Universe}
\author{B.G. Sidharth\\
Centre for Applicable Mathematics \& Computer Sciences\\
B.M. Birla Science Centre, Hyderabad 500 063 (India)}
\date{}
\maketitle
\footnotetext{E-mail:birlasc@hd1.vsnl.net.in}
\begin{abstract}
In this talk, we touch upon the chaotic and fractal aspects of the Universe.
\end{abstract}
\section{Introduction}
It was thought that a subject like Celestial Mechanics belonged to the domain
of deterministic Mechanics\cite{r1}. However studies by Laskar\cite{r2,r3}
subsequently confirmed, that the inner planetary system is chaotic with a
small inverse Lyapunov exponent sheds new light on the topic. This need
not be surprising because the solar system is really a many body system.
Further, it is possible to treat a system with time varying constant of
gravitation as a dynamical system\cite{r4}.
Indeed in certain cosmological schemes the universal constant of gravitation
$G$ varies, with time\cite{r5,r6,r7}.\\
On the other hand there has been an ongoing debate whether the Universe is
homogeneous at the largest scales or if space time is indeed a differentiable manifold
at the smallest scales\cite{r8}.\\
It must be observed that from one point of view, the universe is not a
"continuum" but rather, displays a fractal character. Thus within an atom, the nucleus occupies a very tiny fraction
of the volume, roughly $\sim 10^{-15}$. Then there is a wide gap till we
reach the orbiting electrons. Similarly there are intermolecular distances,
interplanetary, interstellar, intergalactic.... distances which provide
relatively huge gaps. This is not in the spirit of a uniform continuum.\\
We will now argue that this is because, for example the nucleons are bound
together, so also the electrons and the nucleons are bound together, the
atoms in the molecules are bound together... and so on with subsequent gaps, which leads to some
interesting scale dependent consequences, all this in the context of a
Brownian underpinning.
\section{The Fractal Universe}
In \cite{r8} it was argued that we could introduce a "Scaled" Planck Constant
given by
\begin{equation}
h_1 \sim 10^{93}\label{e1}
\end{equation}
for super clusters;
\begin{equation}
h_2 \sim 10^{74}\label{e2}
\end{equation}
for galaxies and
\begin{equation}
h_3 \sim 10^{54}\label{e3}
\end{equation}
for stars.\\
This was directly related to the fact that we have the following Random
Walk relations:
\begin{equation}
R \approx l_1 \sqrt{N_1}\label{e4}
\end{equation}
\begin{equation}
R \approx l_2 \sqrt{N_2}\label{e5}
\end{equation}
\begin{equation}
l_2 \approx l_3 \sqrt{N_3}\label{e6}
\end{equation}
\begin{equation}
R \sim l\sqrt{N}\label{e7}
\end{equation}
where $N_1 \sim 10^6$ is the number of superclusters in the universe,
$l_1 \sim 10^{25}cms$ is a typical supercluster size $N_2 \sim 10^{11}$ is the
number of galaxies in the universe and $l_2  \sim 10^{23}cms$ is the typical size
of a galaxy, $l_3 \sim 1$ light year is a typical distance between stars and
$N_3 \sim 10^{11}$ is the number of stars in a galaxy, $R$ being the radius
of the universe $\sim 10^{28} cms, N \sim 10^{80}$ is the number of elementary
particles, typically pions in the universe and $l$ is the pion Compton
wavelength.\\
The relation (\ref{e7}) was observed nearly a century ago by Weyl and Heddington.
It was shown (Cf.ref.\cite{r7}) that far from being empirical this relation
can be deduced on the basis of the fluctuational creation of particles from a
background Zero Point Field or Quantum Vacuum, in a scheme which leads to a
cosmology consistent also with Dirac's large number coincidences\cite{r9} and
in which the gravitational constant $G$ varies with time,
$$G \propto T^{-1}.$$
From this point
of view the Random Walk character of equation (\ref{e7}) is not accidental, and
this reasoning could be extended to equations (\ref{e4}),(\ref{e5}) and
(\ref{e6}), in the light of equations (\ref{e1}), (\ref{e2}) and (\ref{e3}).
This is against the spirit of deterministic mechanics and it may be mentioned
that it leads to a fractal character\cite{r10}.\\
It may be observed that in all these cases we have a length, the Compton
wavelength or its analogue which defines regions of matter separating
relatively empty spaces.\\
Infact it was also argued in\cite{r8} that these scaled "Compton wavelengths"
and scaled "Planck Constants" arise due to the well known equation of
gravitational orbits,
\begin{equation}
\frac{GM}{L} \sim v^2\label{e8}
\end{equation}
On the other hand equation (\ref{e8}) can be viewed as resulting from the
Virial Theorem\cite{r11}, where the velocity is replaced by the velocity
dispersion.\\
This velocity $v$ would be different at different scales. For example for a black hole
it would be the velocity of light, while for galaxies it is $\sim 10^7 cms$
per second\cite{r12}. It is this circumstance that produces the above scales
leading to fractality.\\
We could go one step further, because we expect that the same effect would
apply to solar type systems: The planets and other objects are bound quite
close to the sun compared to the interstellar distances. Infact we can
verify that this is so for Kuiper Belt objects which have been studied in
the recent past\cite{r13}. In this case a typical size is $\sim 5 km$, the
distances are $\sim 10^{15}cm$, masses are $\sim 10^{19}gms$ their number
is $\sim 10^{10}$ while an application of equation (\ref{e8}) shows that the
velocities are $\sim 10^5 cms$ per second. It can now be shown quite easily
that this defines a scaled Planck Constant $h_4 \sim 10^{34}$.\\
Incidentally from (\ref{e8}) we could easily deduce that the angular momentum
$J$ is given by
\begin{equation}
J \propto M^2\label{e9}
\end{equation}
It is quite remarkable that the equation (\ref{e9}) also applies to elementary
particles and Regge trajectories\cite{r14}. This is a further substantiation
of the rationale for the fractal structure given above in the light of bound
systems separated by relatively large and relatively empty spaces and applies
right upto the level of galaxies\cite{r15}.

\end{document}